\documentclass[useAMS,usenatbib]{mn2e}
\usepackage{txfonts}
\usepackage{graphicx}
\usepackage{textcomp}
\usepackage{natbib}
\usepackage{lscape}
\usepackage{deluxetable}
\usepackage{mathrsfs} 

\newcommand{\um}{\textmu m }
\newcommand{\uu}{\textmu m}

\newcommand{\zirs}{z_{\rm{IRS}} }

\newcommand{\zned}{z_{\rm{NED}} }
\newcommand{\znedalt}{z_{\rm{NEDalt}} }

\newcommand{\zfinal}{z_{\rm{final}} }

\newcommand{\zirseye}{z_{\rm{IRSeye}} }
\newcommand{\zirsalt}{z_{\rm{IRSalt}} }

\bibpunct[]{(}{)}{;}{a}{}{,}

\title[IDEOS redshift paper]{The Infrared Database of Extragalactic Observables from Spitzer I: the redshift catalog}

\author[A. Hern\'an-Caballero et al.]{
Antonio Hern\'an-Caballero,$^1$
Henrik W. W. Spoon,$^{2}$
Vianney Lebouteiller,$^{3}$\newauthor
David S. N. Rupke,$^{4}$
Donald P. Barry,$^{2}$\\
$^{1}$Instituto de F\'isica de Cantabria, CSIC-UC, Avenida de los Castros s/n, 39005, Santander, Spain. E-mail: ahernan@ifca.unican.es\\
$^{2}$Cornell Center for Astrophysics and Planetary Science, Space Sciences Building, Ithaca, NY 14853, USA\\
$^{3}$Laboratoire AIM Paris-Saclay, CEA/IRFU - CNRS/INSU - Universit\'e Paris Diderot, Service d'Astrophysique, B\^at. 709, CEA-Saclay, 91191, Gif-sur-Yvette Cedex, France\\
$^{4}$Department of Physics, Rhodes College, Memphis, TN, 38112, USA}
\begin{document}
\date{Accepted ........ Received ........;}

\pagerange{\pageref{firstpage}--\pageref{lastpage}} \pubyear{2015}

\maketitle

\label{firstpage}

\begin{abstract}

This is the first of a series of papers on the Infrared Database of Extragalactic Observables from Spitzer (IDEOS). In this work we describe the identification of optical counterparts of the infrared sources detected in \textit{Spitzer} Infrared Spectrograph (IRS) observations, and the acquisition and validation of redshifts.
The IDEOS sample includes all the spectra from the Cornell Atlas of Spitzer/IRS Sources (CASSIS) of galaxies beyond the Local Group. Optical counterparts were identified from correlation of the extraction coordinates with the NASA Extragalactic Database (NED). 
To confirm the optical association and validate NED redshifts, we measure redshifts with unprecedented accuracy on the IRS spectra ($\sigma$($\Delta z$/(1+$z$))$\sim$0.0011) by using an improved version of the maximum combined pseudo-likelihood method (MCPL). We perform a multi-stage verification of redshifts that considers alternate NED redshifts, the MCPL redshift, and visual inspection of the IRS spectrum.  
The statistics is as follows: the IDEOS sample contains 3361 galaxies at redshift 0$<$$z$$<$6.42 (mean: 0.48, median: 0.14). We confirm the default NED redshift for 2429 sources and identify 124 with incorrect NED redshifts. We obtain IRS-based redshifts for 568 IDEOS sources without optical spectroscopic redshifts, including 228 with no previous redshift measurements.
We provide the entire IDEOS redshift catalog in machine-readable formats. The catalog condenses our compilation and verification effort, and includes our final evaluation on the most likely redshift for each source, its origin, and reliability estimates.

\end{abstract}


\section{Introduction} 

In the 5.5 years of duration of the \textit{Spitzer Space Telescope} \citep{Werner04} cryogenic mission, the Infrared Spectrograph \citep[IRS;][]{Houck04} performed more than 21,000 observations of $\sim$14,000 distinct targets. 
To further exploit the rich legacy of \textit{Spitzer}, we create the Infrared Database of Extragalactic Observables from Spitzer (IDEOS; Spoon et al. in preparation). The aim of IDEOS is to offer to the community homogeneously measured mid-infrared (mid-IR) spectroscopic observables of the 3575 galaxies beyond the Local Group with observations in the low-resolution ($R$ $\sim$ 60--120) modules of IRS, with spectral coverage between 5.2 and 38.0 \uu. The suite of mid-IR observables that will be available from IDEOS includes restframe continuum fluxes and synthetic photometry, polycyclic aromatic hydrocarbon (PAH) fluxes and equivalent widths, silicate strengths, and mid-IR spectral classifications, among others. IDEOS will also provide optical associations, redshifts, and stitching of the various spectral segments of a target spectrum (even if they were acquired in different observations).
The IDEOS database will provide astronomers with widely varying scientific interests access to diagnostics that were previously available only for limited samples, or available on-the-fly only to expert users.

IDEOS spectra are drawn from the Cornell Atlas of \textit{Spitzer}/IRS Sources\footnote{http://cassis.astro.cornell.edu} \citep[CASSIS;][, hereafter L11]{Lebouteiller11}. CASSIS is an archive of publication-quality spectra containing all the IRS low-resolution staring-mode spectra obtained during the Spitzer mission. CASSIS utilizes an automatic spectral extraction tool based on SMART-AdOpt \citep{Lebouteiller10} that can perform optimal (as well as regular) extractions using a super-sampled point spread function to obtain the best possible signal-to-noise ratio, while still being able to handle blended sources and non-uniform background emission.  

The Spitzer Heritage Archive (SHA)\footnote{http://sha.ipac.caltech.edu} also provides background-subtracted extracted one-dimensional spectra (post-BCD data products) for all \textit{Spitzer/IRS} observations. 
However, CASSIS has several advantages over post-BCD spectra: 
a) CASSIS is an atlas that is fully integrated into a database, including the spectra themselves, thereby allowing sophisticated queries across multiple parameters (e.g., coordinates, detection level, extent, flux density, etc). Local access to the full database is offered by request to users who seek to prune massive datasets.
b) The source extent is derived from a comparison between the source spatial profile and the supersampled point spread function (PSF) profile. From this extent it is evaluated whether the tapered column extraction (aka regular) or the optimal extraction is the most appropriate for the source as far as flux calibration is concerned.
c) Checking for the presence of contaminating sources in the background and selection of by-nod or by-order background subtraction accordingly.
d) Better rogue pixel rejection during image cleaning, using a super-rogue mask rather than just a campaign mask.  
e) CASSIS spectra are defringed. 

A crucial first step in obtaining spectroscopic observables from the extragalactic spectra is to procure accurate and reliable redshifts for the targets. While most low-redshift targets have reliable spectroscopic redshifts derived from other wavelengths (in particular optical), fewer than 50\% of sources at $z$$>$1 have spectroscopic redshifts. In these sources, many of them dusty starbursts, redshifts obtained from the IRS spectrum are at least an order of magnitude more accurate compared to photometric redshifts. In addition, choosing the right optical counterpart of IDEOS sources is not straightforward given the relatively large uncertainty in the coordinates of the extracted source and the faint optical magnitudes of many of them. A redshift estimate obtained directly from the IRS spectrum allows to secure the source identification and confirm the optical redshift, if available.

In \citet[][; hereafter HC12]{Hernan-Caballero12} a method was presented for automatically obtaining accurate redshifts from low-resolution mid-infrared spectra. It finds the redshift of a source as well as an estimate of its reliability by comparing its mid-infrared spectrum with a set of spectral templates. A modified maximum-likelihood algorithm called `maximum combined pseudo-likelihood' (MCPL) searches for local -instead of absolute- minima in the redshift dependency of a pseudo-likelihood function that replaces $\chi^2$. Then, the information on the local minima found with each the template is filtered and combined to produce a more robust result.
The power of the method was demonstrated using a sample of 491 published IRS spectra from the \textit{Spitzer}/IRS Atlas project \citep[ATLAS-IRS;][]{Hernan-Caballero11}.

In this work we use an improved version of the MCPL method that allows to quantify the reliability of MCPL redshift estimates for individual sources, and we apply it to the entire IDEOS sample. We obtain accurate redshifts for 553 IDEOS sources with no previous spectroscopic redshift in the literature. Furthermore, we use a comparison of MCPL redshifts with those in the literature to obtain an independent confirmation of published redshifts, and to identify sources with inconsistent optical associations and/or redshifts. We carefully check those sources individually.  
As a result of this work we produce the IDEOS redshift catalog, which associates an optical counterpart, redshift, and redshift quality information to each of the IRS spectra in the IDEOS sample. The catalog is available online in several machine readable formats\footnote{http://ideos.astro.cornell.edu/redshifts.html}.

The outline of the paper is as follows. Section 2 details the selection of the IDEOS sample. \S3 describes the data reduction and source extraction. \S4 explains the method used for order stitching. \S5 adresses the identification of optical counterparts. \S6 discusses redshifts from the literature. \S7 describes the MCPL method for measurement of redshifts from the IRS spectrum, and our assessment of the accuracy and reliability of MCPL redshifts. \S8 describes the procedure used to verify optical and IRS redshifts. \S9 Describes the IDEOS catalog. Finally, \S10 summarises the main results of the paper.

\section{Sample selection}

The sources included in IDEOS were drawn from the complete set of
$\sim$13000 Spitzer-IRS low-resolution ($R$$\sim$60--120) staring mode observations (corresponding to $\sim$11000 distinct sources)
contained in CASSIS \citep{Lebouteiller11,Lebouteiller15}. This sample therefore excludes
low-resolution spectra obtained in mapping mode, which mostly target very nearby sources.
We used the description of the Spitzer Programs in the SHA to exclude observations of galactic targets. From $\sim$4900 observations of extragalactic sources we included those that fulfilled the following conditions.
i) the galaxy is located beyond the Local Group of galaxies (distance $>$4.5 Mpc).
ii) the galaxy was at least barely detected (CASSIS Detection Level $\geq$1) in one of the spectral orders.\footnote{The detection level quantifies the signal in the spatial profile of the source (see \citet{Lebouteiller10} for details). The typical SNR per pixel for Detection Level = 1 is $\sim$2.}
iii) the pointing was off by no more than one pixel (1.8$"$ in SL and 5.1$"$ in LL) from the nucleus of the galaxy. Spectral segments not fulfilling these criteria were omitted.
iv) the spectrum was not affected by complications in the
background subtraction, such as the presence of emission in the
background image, or the presence of other spurious features such as
a jail bar pattern (see L11).

Several sources in the IDEOS sample have been observed multiple times.
Some observations are outright duplications (mostly calibration
sources), while others contain spectral segments that supplement
previous low-resolution spectra. We have searched the IDEOS sample
for the latter type of incomplete spectra and found 160 sources
for which spectral segments could be merged from two and sometimes
even three observations that were extracted at the same position. The final IDEOS sample contains spectra for 3361 distinct sources.

\section{Data reduction}

The extragalactic sample that constitutes IDEOS is a subset of the Spitzer/IRS atlas CASSIS. We refer to L11 for detailed explanations on the pipeline processing. In this section, we emphasize the aspects of the pipeline that are the most important and relevant for IDEOS, and we also explain the improvements in CASSIS since the seminal paper (L11).

In a nutshell, CASSIS first cleans bad pixels in the exposure images using a rogue pixel mask calculated from prior observations from all campaigns. The exposure images are then combined.\footnote{There was no case in IDEOS when the dispersion in coordinates over the time of the observation was such that images had to be extracted individually instead of being combined and then extracted.} The remaining low-level rogue pixels that were not cleaned and the large-scale background emission is then removed from the science image using two different methods, either removing the observation at the other nod position (\textit{by-nod}), or removing the observations at the two nod positions in the other spectral order (\textit{by-order}). The best background subtraction method is chosen a posteriori based on (1) the presence of contaminating sources in the background image and (2) the signal-to-noise ratio (SNR) of the final spectrum. The spectra are extracted using optimal extraction and tapered column (also referred to as regular) extraction. The optimal extraction uses a super-sampled PSF to fit the spatial profile of the source and is suited for point sources. The tapered column extraction simply integrates the flux in a given window whose width scales with wavelength (reflecting the FWHM increase of the PSF) and with the source spatial extent (as determined from the comparison between the source spatial profile and the super-sampled PSF). The two nod spectra are then defringed, flux calibrated, and combined to produce a single spectrum per observation. The spectral segments corresponding to different spectral orders and different slits are not stitched by default (see \S\ref{sect-stitching}).

Three new CASSIS versions were released after L11. Version 5 (released 2013 March 26) introduced the latest and final calibration for the detector images (S18.18.0), improved the identification of contaminating sources in the potential background images, and improved the background determination for tapered column extractions. The main update in version 6 (released 2014 February 26) concerned the background subtraction for faint sources. The background subtraction method that uses the other spectral order images (\textit{by-order}) removes the large-scale emission assuming the emission in the offset position is identical to the extended emission at the science object position. The residual emission due to this approximation can result in a non-zero baseline which can affect the spectral extraction, both tapered column and optimal extraction, especially in the case of faint sources\footnote{It is impossible to quantify a flux level below which source extraction using \textit{by-order} background subtraction is affected by large-scale residual emission, since it ultimately depends on the relative difference between the source brightness and the residual emission.}. To resolve this issue affecting \textit{by-order} background-subtracted images, CASSIS performs an \textit{additional} large-scale emission subtraction, by calculating and removing the residual \textit{local} spatial continuum before extraction.
CASSIS version 7 (released 2015 June 25), which is the latest version available at the time of our study, built on this improvement by correcting the local spatial continuum also for by-nod background-subtracted images. Version 7 also improves the spectra of tapered column extractions of extended sources in by-nod background-subtracted images.

\section{Order stitching}\label{sect-stitching}

IRS spectra are composed of up to 6 different spectral orders (3 in the Short Low module and 3 in the Long Low). In both modules, the first and second orders overlap in a narrow interval (0.3 and 1.8 \um wide, respectively) which is also sampled by the third `bonus' order spectrum (see Table 1 in L11 for details).

In many spectra, adjacent spectral orders appear shifted in flux from one another, sometimes by a very large factor. There are several causes for that, the most common being the different size of the SL and LL slits, which implies varying amounts of flux lost in resolved sources. 
CASSIS chooses the tapered column extraction method for partially-extended sources because in such cases the flux calibration is more reliable than optimal extraction. The tapered column scales with the source extent in order to sum all the flux falling inside the slit. However, the fraction of light falling outside the slit is not easy to compute since it depends on the source geometry. 

The first and second order spectra of the same module can appear shifted from one another if the background subtraction is not accurate. Also, a slight mispointing perpendicular to the SL slit can have a slitloss effect that is bigger in SL1 compared to SL2, therefore causing a jump between SL1 and SL2.

If left unchecked, flux disparities between spectral orders create unphysical steps in the stitched spectrum, which can cause spurious results when the template fitting routine tries to reproduce them.
Our automated stitching routine removes these steps by scaling the SL spectral orders to LL under the reasonable assumption that the SL spectrum would scale by the same factor if the slit had the same size and orientation of the LL slit.
The stitching algorithm works as follows: the LL2 spectrum, if observed, is taken as reference (that is, remains unchanged). The LL1 spectrum is shifted to match the LL2 flux level, since differences between LL1 and LL2 and usually associated to the background subtraction. SL being a smaller slit, the entire SL spectrum is scaled to get a smooth stitch between SL1 and LL2. Further adjustments can be applied to fine-tune the SL3 to SL1 and SL2 to SL3 stitches if necessary.  We note that this procedure does not ensure an accurate flux calibration throughout the spectrum. 
The scaling factor (or offset) to be applied for each order is obtained as the ratio (or difference) between the integrated flux of the order that works as reference and the one being adjusted in a narrow band within the region where both orders overlap. 

After all orders have been adjusted, the next step is trimming the edges of the spectral orders and stitching them together. The stitching is not performed at a fixed wavelength, but at the wavelength where the spectra for the two overlapping orders intersect. If there are more than one intersections within the integration band, we choose the one where the difference between the slopes of the two spectral orders is smallest. This way we ensure a smooth transition with no steps.

The uncertainty of the scaling factors and offsets is estimated via error propagation from the flux uncertainties (RMS error) for each resolution element. If the one sigma uncertainty in the correction is larger that the correction itself, it is not applied in the automated stitching. Finally, all the spectra are visually inspected and re-stitched interactively if needed. 

\section{Source identification}\label{SourceIDsubsect}

Sources in IDEOS are identified based on the position at which they
were extracted. This information is known to an accuracy that
increases with the number of orders and modules used in the observation.

For an individual module/order/nod observation the position of the
source can be determined to a precision of 1/10 of a pixel in the
cross-dispersion direction (0.2$"$ in SL; 0.5$"$ in LL;
Lebouteiller et al. 2011), while in the dispersion direction
currently no method is implemented to offer
constraints. In practice, though, targets that were acquired using
peak-up imaging\footnote{IRS Handbook} will have the source
well-centered in the slit. For the remaining sources
we will assume the same, but adopt an uncertainty of one
detector pixel (1.8$"$ in SL; 5.1$"$ in LL). 
Since any slit centering offsets will result in slit losses
that are strongest in SL1\footnote{The PSF is largest compared to
the slit width in the first order spectral
segments; and especially at the long-wavelength end.}, a clear
mispointing in the dispersion direction of the SL module can be
readily inferred from a SL1 spectral segment that lies below
the extrapolated SL2 spectrum.

In case spectral segments from both SL and LL have been observed
the position of the source can be much better constrained thanks
to the precise determination (down to 1/10 of a pixel) of the source
position in the almost orthogonal (96 degrees) cross-dispersion
directions of SL and LL. This error box measures 0.2$"$ by 0.5$"$.
In practice we find that the SL and LL spectral segments of a point
source are scattered around the average target position within a
circle of 0.5$"$ radius.

Based on the IRS source position as determined using the method
described above a match is sought with the galaxies contained in
the NASA/IPAC Extragalactic Database (NED). If only one galaxy
is found within the extraction error box, that source is automatically
selected as the counterpart.
If multiple extragalactic sources reside within
the PSF beam a visual inspection of the galaxy population in the
surrounding 7$"$ is conducted. In most of these cases the correct
counterpart is identified by matching the redshift derived from the IRS spectrum ($\zirs$, see \S\ref{sec-zirs}) with the redshift listed by NED ($\zned$) for the nearby galaxies.\footnote{In a few rare cases the correct NED counterpart
was found up to 20$"$ from the extraction position. These galaxies
only had an IRAS source position available for them.}
If spectral features are absent, hard to discern
due to low SNR or due to incomplete spectral coverage, the
source name and/or AOR label supplied by the observer have been
helpful to identify the correct counterpart\footnote{This has
exposed some notable mispointings, where a different source
was observed than intended by the observer.}.
In case NED does not find a galaxy at or near the IRS source
position a photometric catalogue search was conducted. This
generally resulted in a photometric counterpart, mostly
SWIRE2 and SWIRE3 sources\footnote{NED identifiers with prefix SWIRE2 and SWIRE3 correspond to sources from the Data Release 2 and 3, respectively, of the Spitzer Wide Area Infrared Extragalactic Survey \citep[SWIRE;][]{Lonsdale03}}. For most of these, reliable $\zirs$ redshifts
have been determined and assigned using the method described
in this paper.

Several sources in the IDEOS sample have been observed multiple times.
Some observations are outright duplications (mostly calibration
sources), while others contain spectral segments that supplement
previous low-resolution spectra. We have searched the IDEOS sample
for the latter type of incomplete spectra and found 160 sources
for which spectral segments could be merged from two and sometimes
even three observations that were extracted at the same position.

\section{Redshifts from NASA Extragalactic Database}\label{NEDredshifts-section}

In order to obtain valid measurements of infrared spectral features it is essential to obtain reliable redshifts. Our primary source of redshifts is NED.
We query NED servers for redshift measurements corresponding to the unique object identifier that we associated to each AOR following the procedure described in \S\ref{SourceIDsubsect}. Out of 3361 IDEOS sources, 3128 have redshift information in the NED database. 2641 of them have multiple redshift entries in NED, and some of these redshifts are inconsistent with each other (that is, they differ well beyond their uncertainties). We take by default the redshift value preferred by NED ($\zned$), but we also consider the alternative values ($\znedalt$) when comparing with our MCPL redshifts ($\zirs$) at the redshift verification stage (see \S\ref{zfinal-section}).

In $\sim$10\% of IDEOS sources $\zned$ already comes from the IRS spectrum. These are mostly optically faint, high-redshift sources that were observed by several programs targeting (ultra-)luminous infrared galaxies \citep[e.g.][]{Houck05,Weedman06,Yan07,Farrah08,Dasyra09,Hernan-Caballero09,Weedman09}.
While these redshifts are spectroscopic in nature, we consider them separately because they are usually based on the visual identification of PAH features or the 9.8\um silicate absorption band, and their uncertainties ($\Delta$$z$/(1+$z$)$\sim$0.01--0.1) are large compared to both our $\zirs$ measurements and regular optical/radio spectroscopy. 
Therefore, we group NED redshifts into three categories according to their origin: photometric, spectroscopic, or from the IRS spectrum.\footnote{For sources where the origin of the redshift is unclear from NED records we check the original literature reference for the redshift.}
The final statistics for $\zned$ is as follows: 
2655 from optical/radio spectroscopy, 161 photometric, 310 from the IRS spectrum, 2 redshifts of unknown origin, and 228 sources with no redshift in NED (including the 17 sources with no association).

For 964 out of 2641 sources with multiple redshifts in NED the default measurement is spectroscopic and all the alternative values agree within $\vert$$z_{max} - z_{min}$$\vert$/(1+$<$$z$$>$)$<$0.0005. Accordingly, we consider these redshifts to be highly reliable. We use these sources as a control sample to test the accuracy and reliability of MCPL redshift in \S\ref{accuracy-subsec} and \S\ref{reliability-subsec}, respectively. 

There are also many sources with two or more spectroscopic redshift entries in NED that are inconsistent with each other. For 402 sources the difference between extreme values is  $\vert$$z_{max} - z_{min}$$\vert$/(1+$<$$z$$>$)$>$0.005.
Since the typical accuracy of our MCPL redshifts is $\Delta$$z$/(1+$z$)$\sim$0.001 (see \S\ref{accuracy-subsec}), we can use them to decide which one of the redshifts listed by NED is likely to be correct. MCPL redshifts can also verify spectroscopic NED redshifts from a single measurement, and improve the accuracy over photometric and previous IRS-based NED redshifts. The procedure used for this is detailed in \S\ref{zfinal-section}.

\section{Redshifts from the IRS spectrum}\label{sec-zirs}

\subsection{The method}

We obtain redshifts from the IRS spectrum ($\zirs$) for all IDEOS sources using an improved version of the maximum combined pseudo-likelihood (MCPL) algorithm described in HC12.
The most important improvement is the addition of a pre-processing stage where sources are separated according to the slope of their observed-frame spectra. Sources with a very blue MIR continuum (characteristic of the stellar emission that dominates the spectra of early type galaxies) are separated from the main sample and fit with a different set of templates. This eliminates the problem of poor results for ellipticals and radiogalaxies described in HC12.

Very briefly, the algorithm works as follows: it obtains least squares fits for the spectrum with a set of templates {$T_i$} and a grid of redshifts {$z_j$} and computes the $\chi^2$ statistics for every template and redshift, $\chi^2_T(z)$.
A pseudo-likelihood function is defined as $q_T(z)$ $\propto$ 1/$\chi^2_T(z)$, and a `filter' operator zeroes all values of $q_T(z)$ except those that represent local maxima. 
The combined pseudo-likelihood $Q(z)$ is the sum of the filtered $q_T(z)$ over all the templates. Accordingly, $Q(z)$ = 0 at all redshifts except those where one or more of the templates produce a local minimum in $\chi^2_T(z)$. Non-zero values of $Q(z)$ depend both on the number of templates that produce local minima in $\chi^2_T(z)$ at a given redshift and the depth of those minima.

The highest of the peaks in $Q(z)$ marks the MCPL redshift estimate, $\zirs$, which in most cases agrees with the actual redshift of the source. However, if the spectrum is noisy, has very weak spectral features, and/or contains unusual features not found in the templates, it is possible for one or more spurious peaks in $Q(z)$ to be higher than the one corresponding to the true redshift. 
While in such cases $\zirs$ will be wrong, it is still possible to use the remaining peaks in $Q(z)$ to confirm or challenge a redshift measurement obtained by other means, or to choose the right one among several conflicting independent measurements. This is applied in \S\ref{zfinal-section} to the sources with one or more redshifts listed in NED.

For the 241 IDEOS sources with no redshift information in NED as well as the 17 sources with no optical identification, $\zirs$ is the only redshift measurement available. In \S\ref{reliability-subsec} we present an analysis of the properties of $Q(z)$ in the control sample that allows to predict the probability of $\zirs$ being wrong based only on quantities derived from $Q(z)$. 

\subsection{Accuracy of MCPL redshifts}\label{accuracy-subsec}

\begin{figure} 
\begin{center}\hspace{-0.3cm}
\includegraphics[width=8.5cm]{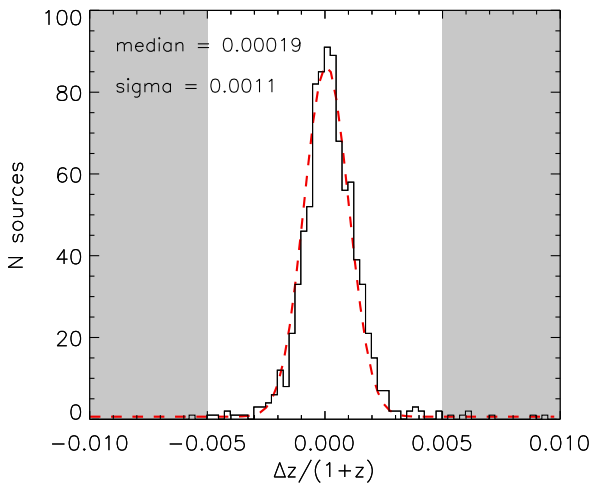}
\end{center}
\caption[]{Distribution of errors in MCPL redshifts for a control sample of 964 IDEOS sources with highly reliable redshifts from NED. The dashed line represents the best-fitting single Gaussian model. The shaded areas indicate redshifts errors $\delta$$>$0.005, which are considered outliers.\label{deltadistrib1}}
\end{figure}

The low resolution modules of the IRS have a resolving power $R$$\sim$100 or $\Delta \lambda$/$\lambda$$\sim$0.01. Even if spectral features can be fitted with sub-pixel precision in high SNR spectra, the uncertainty in the wavelength calibration\footnote{Table 4.3 of the IRS Instrument Handbook lists the following RMS residuals of the wavelength calibration lines: SL1: 0.013 \uu, SL2: 0.008 \uu, LL1: 0.024 \uu, LL2: 0.009 \uu.} ($\Delta \lambda$/$\lambda$ $\sim$ 0.001) ultimately constraints the maximum theoretical redshift accuracy to $\Delta z$/(1+$z$)$\sim$0.001.
 
To evaluate the actual accuracy of $\zirs$ for the IDEOS sources we take the (optical/radio) spectroscopic redshift from NED as the true redshift. A complication arises from the fact that even spectroscopic redshifts are sometimes inaccurate or wrong, as evidenced by the sources with multiple independent redshift determinations in NED (see \S\ref{NEDredshifts-section}).

To minimize the impact of the uncertainty in spectroscopic redshifts in the evaluation of $\zirs$ accuracy, we build a control sample containing only sources with two or more spectroscopic redshift measurements in NED, all of them agreeing within $\Delta z$/(1+$z$)$<$0.0005, a factor of 2 smaller than the maximum theoretical accuracy for IRS redshifts. For the 964 sources that match these criteria ($\sim$1/4 of the total), we assume the actual redshift to be the one that NED chooses by default, $\zned$. 
The error in $\zirs$ is then represented by $d$ = ($\zirs$ - $\zned$)/(1 + $\zned$) and its modulus, $\delta$ = $\vert d \vert$, defines the accuracy of the MCPL redshift. We emphasise that $\delta$ is in strict sense an upper limit to the actual error in $\zirs$ due to the finite uncertainty of $\zned$.

\begin{figure}  
\begin{center}\hspace{-0.3cm}
\includegraphics[width=8.5cm]{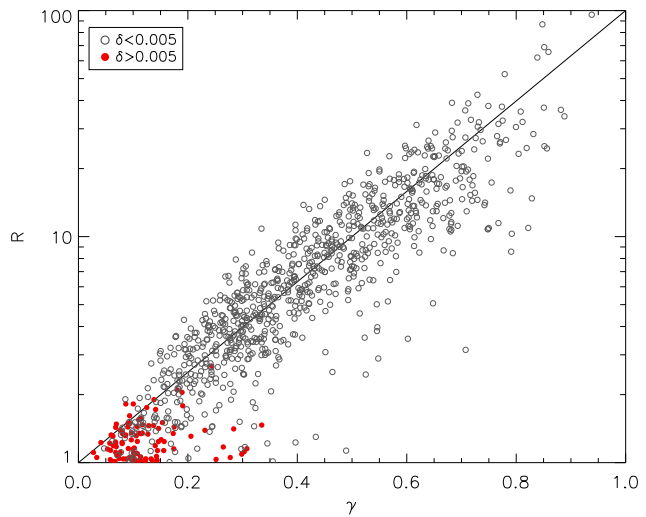}
\end{center}
\caption[]{Correspondence between $R$ and $\gamma$ reliability parameters for the control sample of IDEOS sources with reliable NED redshifts. Open and solid symbols represent, respectively, sources with redshift errors smaller and larger than the $\delta$=0.005 threshold used to define outliers.\label{R-gamma}}
\end{figure}

The distribution of $d$ for the control sample is shown in Figure \ref{deltadistrib1}. Its shape is well approximated by a Gaussian with 1-$\sigma$ dispersion $\sigma_d$=0.0011 and no significant offset. This is comparable to the uncertainty in the wavelength calibration and implies that we achieve a typical accuracy of $\sim$0.1 pixels in the alignment of spectral features between spectra and templates. In comparison, we find that redshifts obtained by visual identification of narrow spectral lines are accurate at the $\sim$0.5--1 pixel level. 

The median value for $\delta$ is 0.0008.  
We conservatively identify as outliers those $\zirs$ with $\delta$$>$0.005, that is, with redshift errors $\gtrsim$4.5$\sigma_d$. Outliers represent 10.9\% of the control sample, and the distribution of their redshift errors is approximately uniform and not consistent with the wings of a Gaussian distribution. This indicates that most outliers are not caused by larger than usual uncertainties in the $\zirs$ estimates, but instead represent a catastrophic error which implies choosing the wrong peak of the $Q(z)$ function. 

The typical redshift error and outlier rate for $\zirs$ in the whole IDEOS sample are probably higher compared to the control sample, because selecting for the more reliable NED redshifts favours bright, low-redshift galaxies, which usually also have high SNR IRS spectra. While the median average SNR per pixel of the sample is 8.9, for 9\% of the sources the average SNR per pixel is $<$2. To quantify how the accuracy of MCPL redshifts depends on the SNR of the spectra we selected the 331 galaxies in the control sample with average SNR per pixel over 20. Then we degraded the SNR per pixel to $\sim$2 by adding Gaussian noise to the spectrum, and run the MCPL algorithm with the degraded spectra.
The results indicate that degrading the SNR per pixel from $>$20 to $\sim$2 increases three-fold the outlier rate, while redshift errors increases two-fold, and no systematic bias is introduced.

\subsection{Reliability of MCPL redshifts}\label{reliability-subsec}

In HC12 it was shown that the numerical value of the absolute minimum of $\chi^2$ (or its MCPL equivalent, the `combined pseudo-likelihood', $Q$) does not provide an indication of the reliability of the $\zirs$, because it depends mostly on the SNR of the spectrum. A noisy spectrum easily obtains good fits (low $\chi^2$) with a broad range of templates and at many different redshifts, while a high SNR spectrum often gets high $\chi^2$ values, even at the correct redshift, because it is unlikely for any of the templates to accurately reproduce all of the features observed in a high SNR spectrum.

A more convenient indicator of the reliability of a redshift solution is obtained by comparing the $\chi^2$ (or $Q$) of the selected solution with the values obtained for alternative solutions. Two reliability indicators of this kind were defined in HC12: $\gamma$ and $R$. $\gamma$ is the value of $Q$($z$) evaluated at the preferred solution and normalised to the integral of $Q$($z$) over the entire redshift search range, while $R$ is the ratio between the first and second highest values of $Q$($z$). The accuracy and reliability of MCPL redshifts were both shown to increase monotonically with $\gamma$ and $R$. In addition, it was shown that while the fraction of outliers depends on the mid-infrared spectral class (redshifts are more reliable in PAH-dominated IRS spectra compared to continuum dominated spectra), it is largely independent of the mid-infrared classification among sources with comparable values of $\gamma$.

Figure \ref{R-gamma} shows the distribution of $R$ versus $\gamma$ for the control sample of IDEOS sources with reliable NED redshifts. 
Sources with accurate $\zirs$ redshifts ($\delta$$<$0.005) form a diagonal sequence that surrounds the line defined by $\log_{\rm{10}}$(R) =  2$\gamma$, while outliers ($\delta$$>$0.005) occupy only the lower left end of this sequence, with many of them very close to R=1, which indicates another redshift solution existed with a nearly identical $Q$($z$) value. 
All the outliers with $\gamma$$>$0.2 correspond to spectra with a very strong 10\um silicate feature, either in emission or in absorption. This somehow degrades the accuracy of the $\zirs$ value (they have 0.005$<\delta<$0.05) probably due to the diversity of shapes that the 10\um silicate feature takes in observed spectra.  

We demonstrate in appendix \ref{reliability_appendix} that the redshift quality (as measured by the typical redshift error and the outlier rate) increases monotonically along the diagonal sequence, but is largely insensitive to a displacement orthogonal to the sequence. Therefore, a single quantity, $A$ = $\gamma$ + $\log_{\rm{10}}$(R)/2, condenses our knowledge of the redshift quality of $\zirs$ prior to a comparison with NED redshifts.

For the control sample we fit with simple parametric models the frequency of outliers, $f_{0.005}$, and the standard deviation of redshift errors after excluding outliers, $\sigma_d$, and  as functions of $A$. By assuming the same relations hold for the entire IDEOS sample, we can predict the outlier probability and uncertainty of individual $\zirs$ measurements:

\begin{equation}
P(\Delta z/(1+z) > 0.005) =  f_{0.005}(A) = 10^{-4.70 A^{1.53}}
\end{equation}
\begin{equation}
\sigma(\Delta z/(1+z)) = \sigma_d(A) = 0.00142 - 0.00075 A
\end{equation}

\begin{figure} 
\begin{center}\hspace{-0.3cm}
\includegraphics[width=8.5cm]{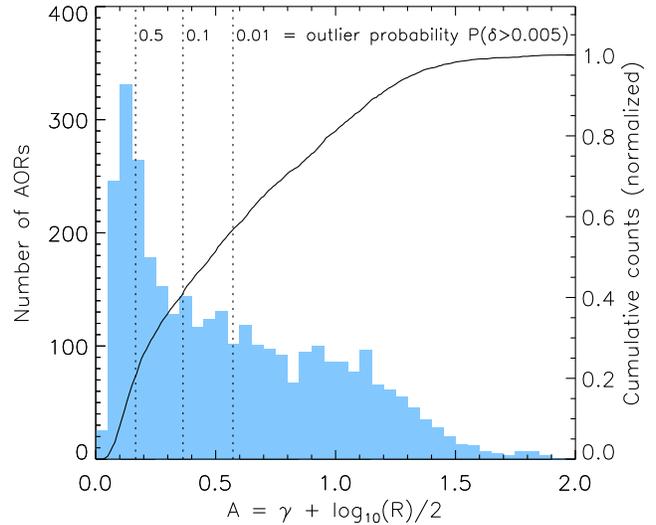}
\end{center}
\caption[]{Distribution of values for the reliability parameter $A$ in the IDEOS sample (histogram) and normalized cumulative distribution (solid line). The dotted lines mark the values of $A$ corresponding to outlier probabilities of 0.5, 0.1, and 0.01.\label{Ahistog}}
\end{figure}

Columns 13 and 14 in Table \ref{redshift-table} list the outlier probabilities at the $\delta$$<$0.005 and $\delta$$<$0.05 accuracy levels for the individual sources in IDEOS. 
The distribution of $A$ for the whole sample (Figure \ref{Ahistog}) implies $\zirs$ is accurate within $\delta$$<$0.005 with $>$90\% confidence for 58\% of the sample, and $>$99\% confidence for 42\%.
However 20\% of the sources have $A$$<$0.17, which implies a probability $>$50\% for the error in $\zirs$ to be larger than $\Delta z$/(1+$z$) = 0.005. These highly unreliable $\zirs$ measurements are mostly useless on their own, but can still help confirm the optical redshift from NED or choose among several inconsistent values. 

\section{Redshift verification}\label{zfinal-section}

\begin{figure*} 
\begin{center}\hspace{-0.3cm}
\includegraphics[width=8.2cm,height=7.5cm]{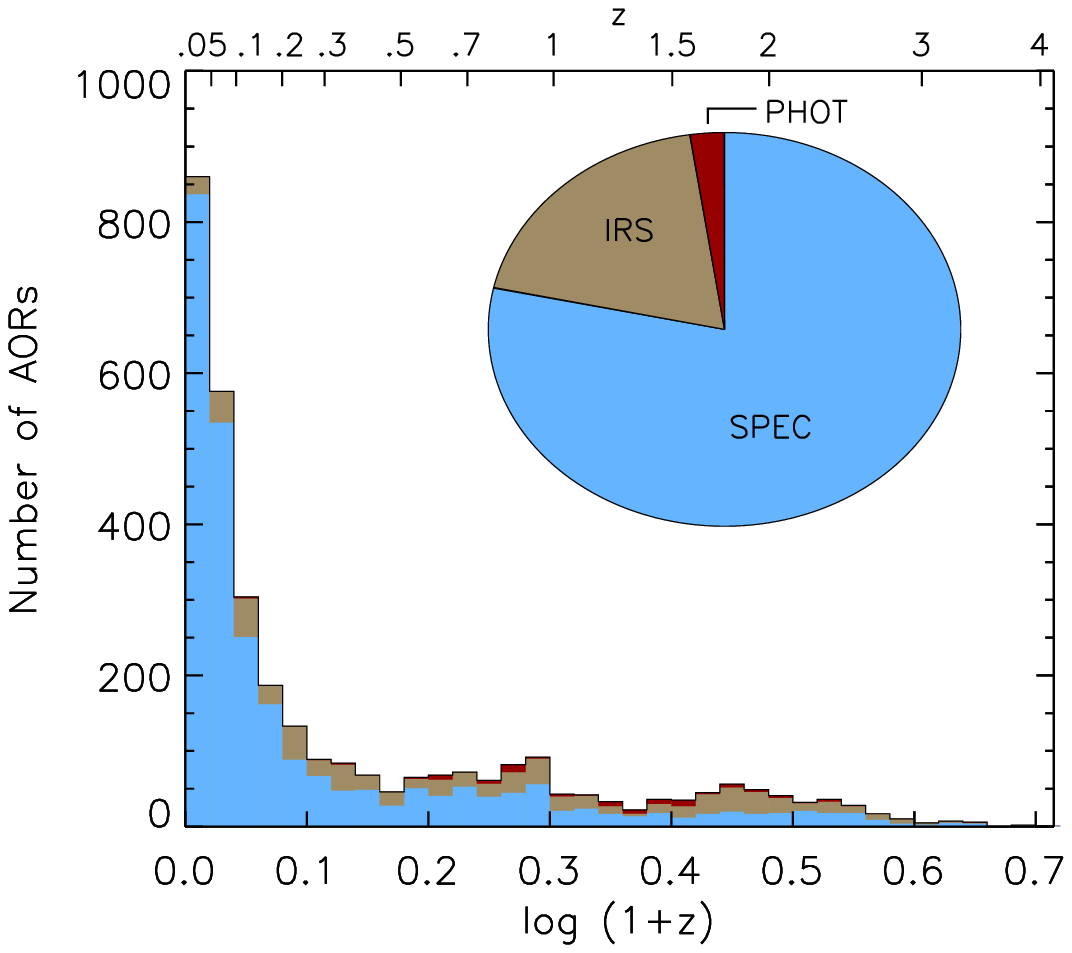}
\hfill
\includegraphics[width=8.2cm,height=7.5cm]{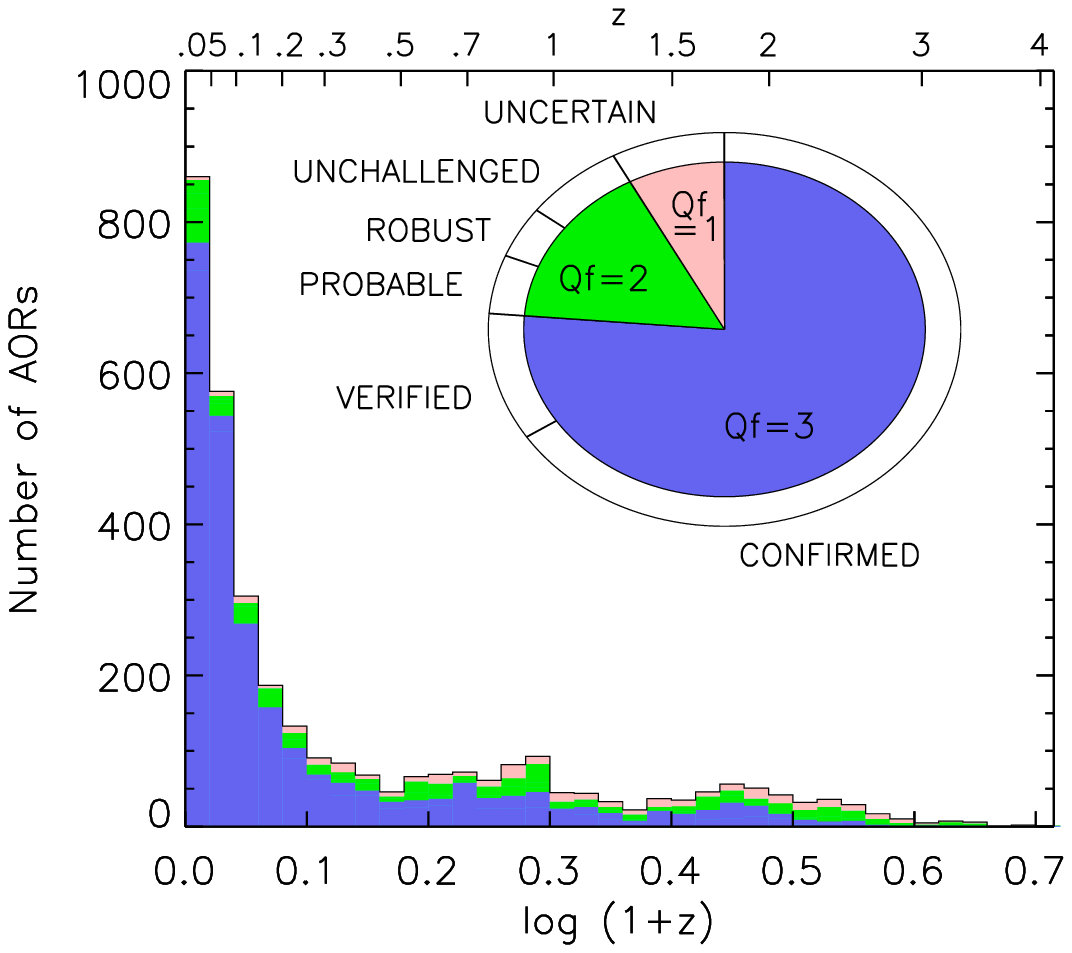}
\end{center}
\caption[]{Redshift distribution of the IDEOS sample, with colour coding for the origin of $\zfinal$ (left) or its verification status (right). The pie diagrams represent the fraction that each class contributes to the total IDEOS sample. The 5 IDEOS sources with no redshifts are not included.\label{zdistrib}}
\end{figure*}

In this section we provide the details on the procedure used to verify the redshifts of the IDEOS sources and choose the best redshift value ($\zfinal$) for each source. We use different names to identify the verification status of the redshifts, which depends on the results of several tests, as explained below. The verification states that we use are (in order of decreasing confidence): `confirmed', `verified', `probable', `robust', `unchallenged', and `uncertain'.
 
The large majority of IDEOS sources (3104) have both NED and MCPL redshifts. If the default NED redshift and the $\zirs$ agree within $\delta$$<$0.005, we consider the redshift to be `confirmed'. This is the case for 2156 sources (64\% of the IDEOS sample). The margin is narrow enough for a spurious agreement to be highly unlikely (see \S\ref{sec_prandommatch} for details). When the redshift from NED is spectroscopic, we take $\zfinal$ = $\zned$ because we assume it has higher accuracy than the expected $\delta$$\sim$0.0011 uncertainty of $\zirs$. However, if $\zned$ is photometric or derived from the IRS spectrum, we take $\zirs$ instead.

When $\zned$ and $\zirs$ do not agree, we consider alternate $\zirs$ solutions from secondary peaks in the $Q$($z$) function. If we find that the $n^{th}$ peak in $Q$($z$) is within $\delta$$<$0.005 of $\zned$, we calculate the probability of a random match (see \S\ref{sec_prandommatch} for details). We consider the redshift `confirmed' if the probability of a random match is $<$1\% (167 cases), or `robust' if it is between 1\% and 5\% (90 cases).

For the remaining 691 sources, where no agreement between $\zned$ and $\zirs$ is found, we resort to visual inspection of the IRS spectrum.  
We are able to identify spectral features (emission lines, PAH bands, or the $\sim$10\um silicate band) conclusively in 238 sources, for which we obtain a visual redshift ($\zirseye$). For these sources, we select as the final redshift the one that best agrees with $\zirseye$. This is the default $\zned$ in 107 cases, one of the alternate NED redshifts in another 5, and $\zirs$ in 113 sources.
All these redshifts consistent with the visual inspection are classified as `verified'.

There are also 25 sources where $\zirseye$ does not agree with either $\zned$, $\znedalt$, or $\zirs$. However, in nearly all of them we find one or more secondary peaks in $Q$($z$) consistent with $\zirseye$. If the peak is strong and isolated (in 1/2 of cases it is the 2$^{nd}$ or 3$^{rd}$ highest peak) we choose this $\zirsalt$ value as the final redshift and also classify as `verified'. However, if there is a cluster of weaker peaks in $Q$($z$)\footnote{For sources with no PAH bands or emission lines, it is common to find a cluster of peaks in $Q$($z$) at the visual redshift. The larger than usual uncertainty of visual redshifts based on the silicate feature alone makes it very difficult to identify the correct peak.} we choose $\zirseye$ as the final redshift and classify as `uncertain' to reflect the larger uncertainty in the redshift.

In 453 out of 691 sources where $\zned$ and $\zirs$ disagree, the visual inspection was inconclusive due to a very noisy or featureless IRS spectrum. Since the reliability of $\zirs$ for these sources is low (their average outlier probability is $\langle P(A) \rangle$=0.58), we assume $\zned$ is more likely to be correct. We take $\zfinal$ = $\zned$ and classify them according to the reliability of the $\zned$: sources in the control sample of highly reliable NED redshifts (see \S\ref{NEDredshifts-section}) are classified as `probable', other spectroscopic $\zned$ are classified as `unchallenged', and photometric NED redshifts as well as those NED redshifts derived from the IRS spectrum are classified `uncertain'.

Finally, for the 228 IDEOS sources with $\zirs$ but no redshift in NED we take $\zfinal$ = $\zirs$, and classify as `probable' if the outlier probability is $<$1\%, `robust' if 1\%$<$$p$$<$5\%, or `uncertain' if $p>$5\%. 

Figure \ref{zdistrib} shows the distribution of $\zfinal$ values broken up by redshift type and verification status. $\zfinal$ is spectroscopic in 2620 sources, photometric in 78, and derived from the IRS spectrum in 644. The verification status is `confirmed' for 2209 sources, `verified' in 352, `probable' in 162, `robust' in 142, `unchallenged' in 230, and `uncertain' in 261. For 5 sources we could not obtain any redshifts (therefore, no redshift-dependent observables can be measured).

We assign numerical quality flags ($Qf$) to the redshifts according to their verification status. Sources with no redshift have $Qf$=0, while `uncertain' redshifts have $Qf$=1. Reliable redshifts (sources with `unchallenged', `probable', or `robust' status) have $Qf$=2, and $Qf$=3 is reserved for the very reliable `confirmed' and `verified' redshifts. 

We obtain reliable ($Qf$$>$1) redshifts for 3095 out of 3361 sources in the IDEOS sample.
The redshift distribution peaks at $z$$\sim$0 and has an extended tail up to $z$$\sim$3, with only 23 sources at higher redshift. A secondary peak at $z$$\sim$0.9 contains an unusually large proportion of unchallenged and uncertain redshifts. This is a consequence of a single Spitzer program, \#50196 (P.I. Rieke) which observed 57 optically selected QSOs at $z$$\sim$0.8 only in the SL1 module. 
It is also noteworthy the high fraction of sources with $\zfinal$ from IRS at $z$$>$1 ($\sim$40\%). This is due to the large number of optically faint sources observed by programs targeting mid-IR selected infrared-luminous galaxies.

While we do not claim the IDEOS sample to be representative of the more than 5 million galaxies with redshifts in the NED database, their redshift distributions are not very different,\footnote{see http://ned.ipac.caltech.edu/help/ned\_holdings.html for the distribution of all NED redshifts} so we can obtain some statistics that are useful to estimate the overall quality of NED redshifts.
Many of the sources in the IDEOS sample have more than one redshift measurement in the literature. NED lists all of them. While they usually differ only very slightly, there are 402 IDEOS sources for which the difference is large enough to be able to distinguish the correct one using the IRS spectrum ($z_{max}$ - $z_{min}$ $>$ 0.005(1+$<z>$). We find that only in 5 out of 402 sources the IRS spectrum favours one of the alternate NED redshifts instead of the default one. In some cases the alternate redshifts are photometric while the default one is spectroscopic, making the selection straightforward, but in more that half of the cases two or more of the incompatible redshifts are spectroscopic. Therefore, this very small rate of errors reflects an outstanding work of redshift validation by the NED team, that chose the right measurement in $\sim$99\% of cases. 
There are, however, 124 IDEOS sources for which the default redshift from NED does not agree within $\delta<$0.005 with the $\zfinal$ that we verified through visual inspection of the IRS spectrum. Unsurprisingly, these are mostly high redshift sources whose NED redshifts are photometric or based on the IRS spectrum. 70\% of them have redshift errors 0.005$<$$\delta$$<$0.05, that are consistent with the typical uncertainties of photometric redshifts and redshifts derived from the visual identification of the broad 9.7\um silicate feature in the IRS spectrum.
Only 23 of the 124 wrong or inaccurate NED redshifts are spectroscopic (including the 5 where one of the alternate NED redshifts is correct). 
We emphasize that in those cases our analysis implies that the $\zned$ for the optical source that we associate to the IRS spectrum is wrong. An alternative interpretation would be that the IRS spectrum was associated to the wrong optical source. While this is possible in principle, we performed several checks to reduce its likelihood. All cases of multiple optical counterparts within the 2-$\sigma$ error circles of the IRS source extraction coordinates were carefully inspected, and if any of the candidates had $\zned$ compatible with $\zirs$ it was considered the true optical counterpart. In many cases the observer used the name of the intended target in the AOR label or in the object name field. This also serves as confirmation. All this information is available to the IDEOS user when they perform a single source search.

\section{The IDEOS redshift catalog}\label{sect-catalog}

Table \ref{redshift-table} shows an excerpt of the IDEOS redshift catalog. A machine readable version of the full table is available online\footnote{http://ideos.astro.cornell.edu/redshifts.html}.  
The first column indicates the default NED name for the optical counterpart of the infrared source observed by Spitzer. The second column indicates the astronomical observation request (AOR) identifier, with the cluster identifier as a subscript (the cluster ID is zero for non clustered observations). The 3$^{rd}$ and 4$^{th}$ columns indicate the final redshift value and its origin (zNED for default NED redshifts, zNEDalt for alternate NED redshifts, zIRS for MCPL redshifts, and zIRSeye for redshifts from visual inspection of the IRS spectrum). The 5$^{th}$, and 6$^{th}$ columns contain a numerical quality flag [0-3] for $\zfinal$ (higher is better) and the verification status for the redshift. Columns 7 and 8 indicate the default NED redshift and its type (spectroscopic, photometric, or from the IRS spectrum). Column 9 indicates the MCPL redshift solution, $\zirs$, while \#10, \#11, and \#12 represent the $\gamma$, $R$, and $A$ reliability parameters for the MCPL solution, respectively. Columns 13 and 14 represent the probability of $\zirs$ being an outlier at the accuracy thresholds $D$=0.005 and $D$=0.05, respectively, based on the value of $A$ alone. Finally, columns 15 and 16 contain additional information for the cases where the default $\zned$ and $\zirs$ redshifts do not agree but some of the alternate solutions do. Column 15 indicates the redshift corresponding to the peak of $Q$($z$) that agrees with one of the NED redshifts, and column 16 indicates its rank when sorting by decreasing values of $\gamma$. 

\section{Summary}

This is the first of a series of papers that describes the Infrared Database of Extragalactic Observables with Spitzer (IDEOS). In this work we described the procedure followed to securely identify the optical counterparts of the infrared sources detected in \textit{Spitzer}/IRS observations, and the acquisition and validation of redshifts.

The IDEOS sample includes all the spectra from CASSIS of extragalactic sources beyond the Local Group. We use the latest version of CASSIS (v7.0) and a careful stitching procedure to maximize the data quality. Optical counterparts were identified from correlation with the NED database, taking into account the uncertainty in the extraction coordinates, PSF beam size, and the redshifts of the candidates.

Published redshifts for the optical counterparts were retrieved from NED. In addition, redshifts were measured on the Spitzer/IRS spectra using a refined version of the MCPL method described in HC12. We compare the redshifts from NED and from the IRS spectra to validate both the source associations and the redshifts. 

Using a subsample of 964 sources with highly reliable spectroscopic NED redshifts, we find that MCPL redshifts have a typical accuracy of $\sigma$($\Delta z$/(1+z)) = 0.0011, with a 10.9\% rate of outliers. We find the rate of outliers to be strongly dependent on the quality parameter $A$, which quantifies the confidence in the MCPL redshift based on analysis of the IRS spectrum alone. This allows us to identify sources with very reliable MCPL redshifts (42\% of the sample has outlier probability $<$1\%), as well as those highly uncertain (20\% has $>$50\% probability of outlier). 

We perform an automated comparison of NED and MCPL redshifts that takes into account the origin (photometric, spectroscopic) of the NED redshifts as well as the quality of MCPL redshifts. NED and MCPL redshifts are found to agree in 68\% of cases. For the remainder, we resort to a careful multi-stage verification that involves considering alternate NED redshifts and secondary solutions from the MCPL algorithm, as well as visual inspection of the IRS spectrum. We find among them 5 sources for which NED lists the correct redshift but not as the default one, and another 119 sources where $\zned$ is inaccurate or wrong (mostly photometric redshifts or from the IRS spectrum, but 18 of them are optical spectroscopic). We obtain accurate IRS-based redshifts for 568 IDEOS sources without optical spectroscopic redshifts, including 228 IDEOS sources for which no previous redshift measurements were available.

We provide the entire IDEOS redshift catalog in machine-readable formats. The catalog condenses our compilation and verification effort, and includes our final evaluation on the most likely redshift for each source and its origin and reliability, among others.

\section*{Acknowledgements}

We thank the anonymous referee for useful comments that helped to improve this paper.
A.H.-C. acknowledges funding by the Universidad de Cantabria Augusto Gonz\'alez Linares programme and the Spanish Ministry of Economy and Competitiveness under grant AYA2012-31447, which is partly funded by the FEDER program.
H.S. acknowledges funding by NASA ROSES 2012 under grant NNX13AE69G.
This work is based on observations made with the \textit{Spitzer Space
Telescope}, which is operated by the Jet Propulsion Laboratory, Caltech
under NASA contract 1407.
The Cornell Atlas of Spitzer/IRS Sources (CASSIS) is a product of the Infrared Science Center at Cornell University, supported by NASA and JPL.
This research has made use of the NASA/IPAC Extragalactic Database (NED) which is operated by the Jet Propulsion Laboratory, California Institute of Technology, under contract with the National Aeronautics and Space Administration.

\appendix

\section{Reliability predictions for individual MCPL redshifts}\label{reliability_appendix}

\begin{figure} 
\begin{center}\hspace{-0.3cm}
\includegraphics[width=8.5cm]{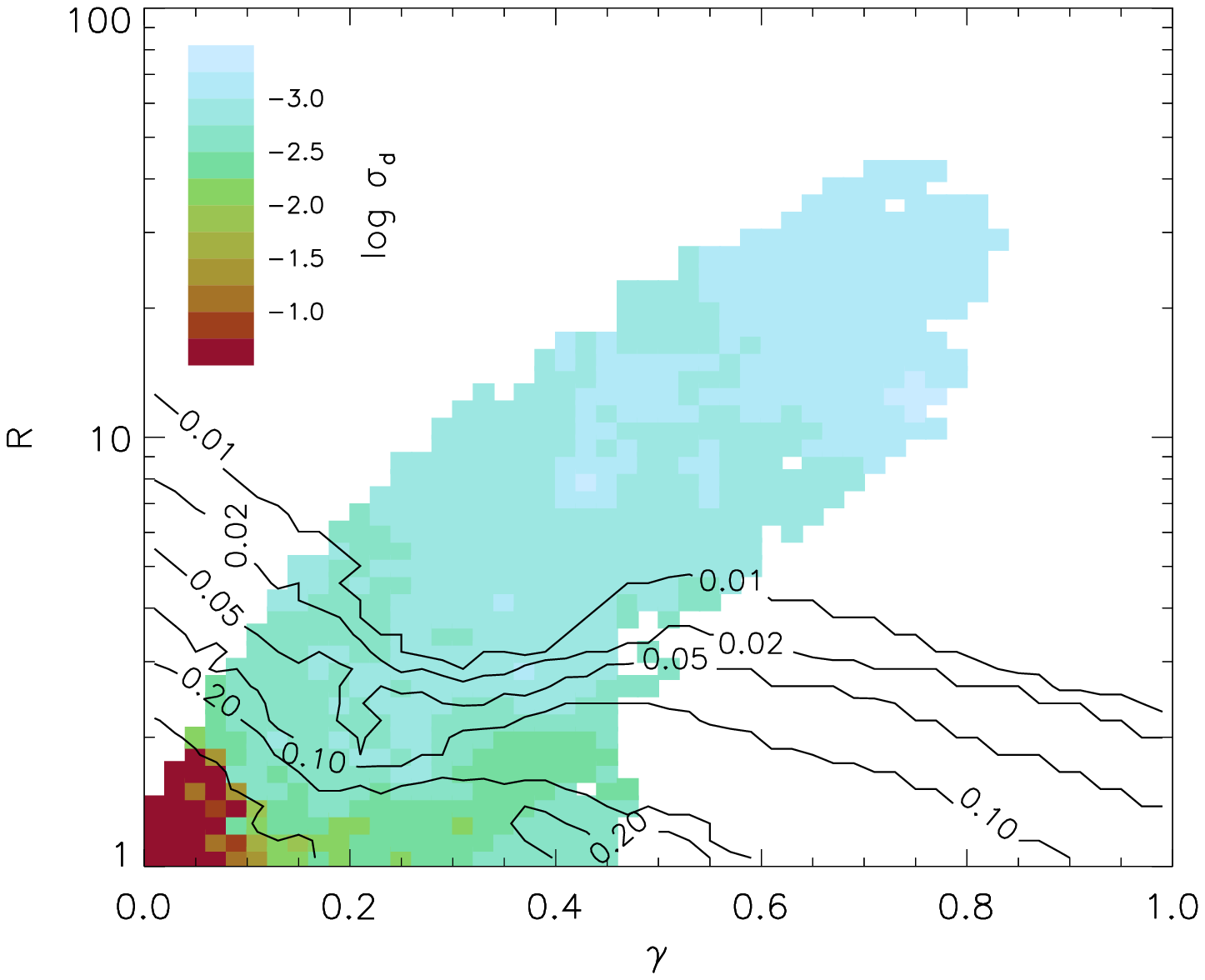}
\end{center}
\caption[]{Distribution of the typical redshift accuracy ($\sigma_d$, color map) and outlier fraction ($f_{0.005}$, contours) for the extended control sample as a function of the reliability parameters $\gamma$ and $R$. See text for details.\label{reliabilitymap}}
\end{figure}

\begin{figure} 
\begin{center}\hspace{-0.3cm}
\includegraphics[width=8.5cm]{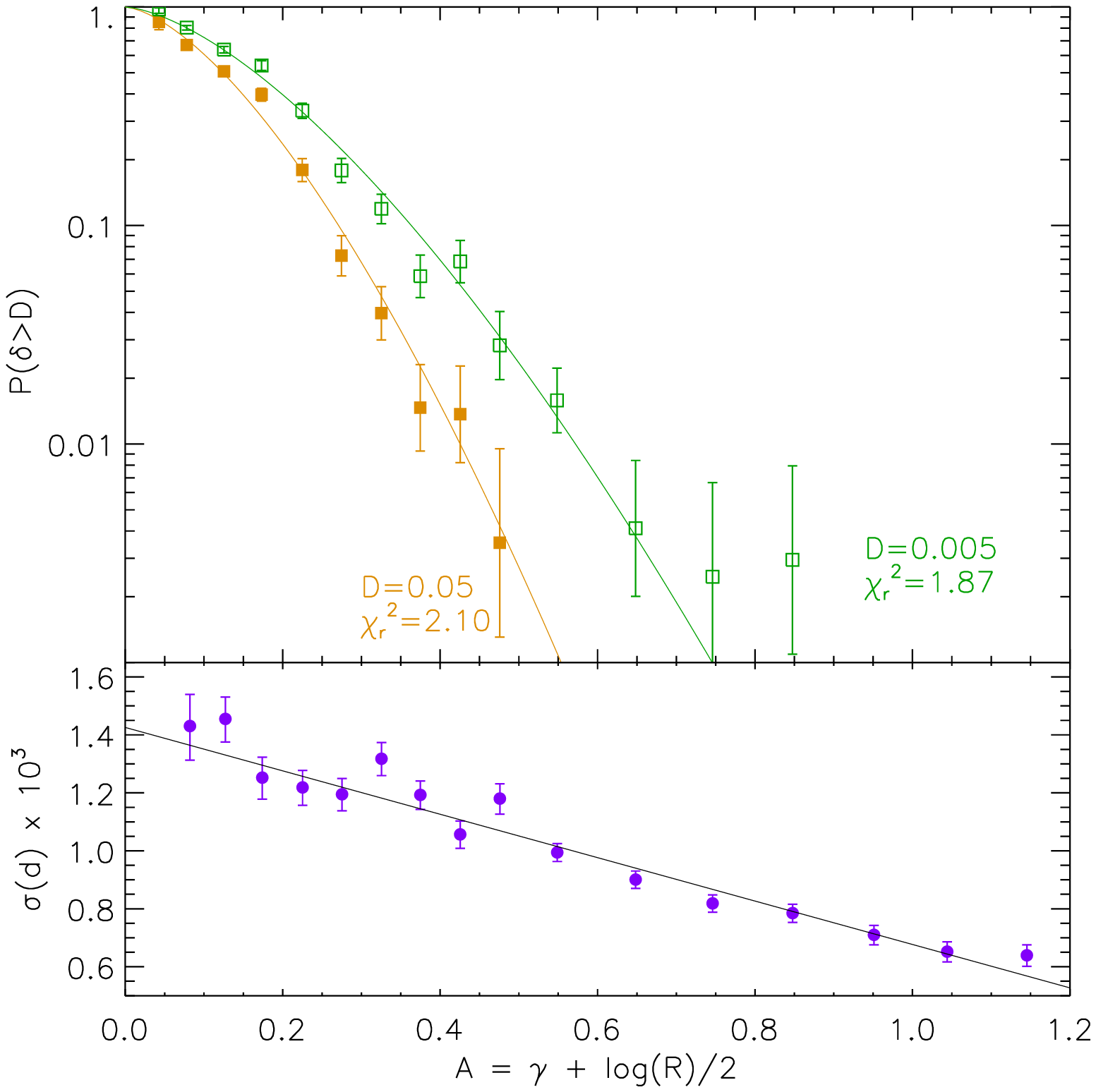}
\end{center}
\caption[]{Top: frequency of outliers in the extended control sample as a function of the reliability parameter $A$ for the accuracy thresholds $D$ = 0.05 (solid symbols) and $D$ =  0.005 (open symbols). Each point represents the fraction of outliers among sources in bins of $A$ with width 0.05 for $A$$<$0.5 and 0.1 for $A$$>$0.5. Error bars represent 68\% confidence intervals calculated with the Wilson formula for binomial distributions. Solid lines represent the best-fitting model of the form: $\log_{\rm{10}}$$P$ = -$\alpha A^\beta$. Bottom: Standard deviation of redshift errors (excluding outliers) as a function of the reliability parameter $A$. Error bars represent 68\% confidence intervals. The solid line represents the best fitting linear model.\label{Poutlier}}
\end{figure}

In this appendix we use the control sample of sources with very reliable NED redshifts to
quantify how the average accuracy and frequency of outliers for $\zirs$ measurements depend on the reliability parameters $\gamma$ and $R$ associated to the $\zirs$ solution. We use this to model the expected accuracy of $\zirs$ and probability of outliers for the remaining sources in the IDEOS sample.

To improve the statistics, we enlarge the control sample up to five thousand spectra by producing multiple versions of each spectrum in the control sample with varying amounts of noise added. $Q$($z$), $\zirs$, and the reliability parameters $\gamma$ and $R$ are re-computed for each of these degraded versions of the original spectra.

We represent the typical redshift accuracy for a given combination of $\gamma$ and R with the standard deviation $\sigma_d$($\gamma$,R) of the redshift errors among sources with comparable $\gamma$ and R values. The frequency of outliers at the $\delta$$>$0.005 level, $f_{0.005}$($\gamma$,R), is the fraction of sources in the control sample with $\delta$$>$0.005 as a function of $\gamma$ and $R$. 

Figure \ref{reliabilitymap} shows the dependency of $\sigma_d$ and $f_{0.005}$ with $\gamma$ and $R$ for the extended control sample. $\sigma_d$ is computed in the 30 sources closest to the center of each tile in the colour map, while $f_{0.005}$ is calculated for the 200 closest sources, in order to reduce Poisson noise in regions with low $f_{0.005}$.
The dispersion in redshift errors, $\sigma_d$, decreases along the diagonal sequence, and is nearly insensitive to a displacement orthogonal to the sequence. This effect (which is stronger closer to the origin of coordinates) implies that $\sigma_d$ decreases with R for fixed $\gamma$ and vice versa. 
The same trend applies to the frequency of outliers at the $\delta$$>$0.005 level, $f_{0.005}$, shown as contours in Figure \ref{reliabilitymap}.
This indicates that a combination of the two reliability indicators, $A$ = $\gamma$ + $\log_{\rm{10}}$(R)/2, condenses most of the information provided by the pair ($\gamma$,R). $A$ also offers stronger correlation with $\sigma_d$ and $f_{0.005}$ compared to $\gamma$ or $R$ alone. Because of this, we choose to model the redshift errors and outlier fraction as a function of $A$ alone.

The top panel in Figure \ref{Poutlier} represents the frequency of outliers in the extended control sample as a function of $A$, $f_{\rm{D}}$($A$), at the accuracy thresholds $D$ = 0.05 and 0.005. The fraction of outliers decreases monotonically and steeply with increasing $A$, and the rate of decrease is faster for higher values of the threshold $D$. 

We model the frequency of outliers with a power-law, $f_{\rm{D}}$($A$) = $\alpha A^\beta$, where $\alpha$ and $\beta$ are adjustable parameters. The best-fitting models obtain $\chi^2_\nu$$\sim$2, indicating the model is a reasonably description of the dependency of the outlier rate with $A$ and $D$. 
Given a sufficiently large control sample, we can approximate the outlier probability as a function of $A$ by the frequency of outliers among sources in the control sample that have the same value of $A$: P($\delta$$>$D $\vert${ } $A$) $\approx$ $f_{\rm{D}}$($A$). For $D$=0.005, this means P($\delta$$>$0.005) = 10$^{-4.7 A^{1.53}}$, which implies that the probability of $\zirs$ being accurate within $\delta$$<$0.005 is $\sim$90\% ($\sim$99\%) for $A$=0.4 (0.6). 

The bottom panel in Figure \ref{Poutlier} shows the dependency of the typical redshift accuracy (as measured by the standard deviation of redshift errors, $\sigma_d$) as a function of $A$, after excluding outliers. There is a steady decrease of $\sigma_d$ with $A$ that is consistent with a linear relation. The best fit is given by $\sigma_d$ = 0.00142\,-\,0.00075\,$A$ ($\chi^2_\nu$ = 1.75). For $A$$>$0.6 the observed dispersion in redshift errors becomes smaller than the uncertainty in the wavelength calibration of IRS spectra. We interpret this result as a sign that our redshift finding method is largely insensitive to uncertainty in the \textit{absolute} wavelength calibration, which mostly cancels out when IRS spectra are used both as data and templates. Instead, our method would be affected only by relative variations in the wavelength calibration between observations, which are expected to be much smaller.

\section{Estimating the probability of a spurious match between $\zned$ and $\zirs$}\label{sec_prandommatch}

Spurious peaks in the Q($z$) function (that is, those not corresponding to the actual redshift of the source) are expected to distribute randomly in the redshift search range (see HC12). Accordingly, the probability of finding $\zirs$ within $\delta$$<$$D$ from $\zned$ is:

\begin{equation}
P(D) = \frac{2D}{\ln(1+z_{max})}
\end{equation}
where $z_{\rm{max}}$ is the upper limit of the redshift search range. For $D$=0.005 and $z_{\rm{max}}$ = 4, this implies a probability $<$1\% for a spurious confirmation with the primary $\zirs$ solution.

In some cases the peak of $Q$($z$) corresponding to the actual redshift of the source is not the strongest one. This is often the case in noisy spectra and those with very weak or unusual spectral features. If we consider the $n$ highest peaks in $Q$($z$), the probability of finding any of them within $\delta$$<$$D$ of $\zned$ by chance is then:

\begin{equation}\label{EQ_Prob_n}
P_n(D) = 1 - (1 - P(D))^n
\end{equation}

The calculation can be generalised to the case with multiple $z_{NED,j}$ regardless of whether they are mutually (in)compatible if we assume that all the $z_{NED,j}$ are equally likely a priory. In sources with both photometric and spectroscopic $\zned$ measurements, we consider only the latter. 

Let $r_{\rm{min}}$ be the smallest distance in $\ln$(1+$z$) units between any of the $z_{NED,j}$ and any of the $n$ strongest peaks in $Q$($z$). The set containing every possible redshift within $r_{\rm{min}}$ of any of the $z_{NED,j}$ is defined by the union of open intervals:

\begin{equation}\label{EQ_union}
S = \bigcup_{j=1}^N{ } (\ln (1+z_{NED,j}) - r_{min},{} \ln (1+z_{NED,j}) + r_{min})
\end{equation}

Then the probability of finding a random value in the [0,$\ln$(1+$z_{\rm{max}}$)] interval inside this set is:

\begin{equation}
P(S) = \frac{L(S)}{\ln(1+z_{\rm{max}})}
\end{equation}

\noindent where L($S$) is the length (Lebesgue measure) of $S$. If the $N$ $z_{NED,j}$ are all at a distance greater than 2$r_{\rm{min}}$ from each other, then all the intervals in Eq. \ref{EQ_union} are disjoint and therefore L($S$) = 2$N$$r_{\rm{min}}$; otherwise, the overlap among intervals implies L($S$) $<$ 2$N$$r_{\rm{min}}$.

Finally, by substituting P($D$) by P($S$) in Eq. \ref{EQ_Prob_n} we obtain the probability of a random match for any of the $n$ strongest peaks in $Q$($z$) with any of the $z_{NED,j}$: 
\boldmath
\begin{equation}
P_n(S) = 1 - \Big{(}1 - \frac{L(S)}{\ln(1+z_{\rm{max}})}\Big{)}^n
\end{equation}

\onecolumn

\begin{landscape}
\begin{deluxetable}{l l l l l l l l l l r l l l l r} 
\tabletypesize{\small}
\tablecaption{IDEOS redshift catalog\label{redshift-table}}
\tablewidth{0pt}
\tablehead{\colhead{NED name} & \colhead{AOR ID} & \colhead{$z_{final}$} & 
\colhead{source} & \colhead{Qf} & \colhead{status} & \colhead{$\zned$} & \colhead{ztype} & \colhead{$\zirs$} & 
\colhead{$\gamma$} & \colhead{$R$} & \colhead{$A$} & \colhead{P$_{0.005}$($A$)} &
\colhead{P$_{0.05}$($A$)} & \colhead{$z_{IRSalt}$} & 
\colhead{order} }
\startdata
[HB89] 2227-394                &  14759424\_0 &  3.43800  & zNED    & 2 & UNCHALLENGED &  3.43800  &  SPEC &  0.36956  &    0.047 &  1.02   &    0.051 &    0.8885 &    0.8280 &          &   \\
5MUSES 027                     &  24180224\_5 &  1.16402  & zIRSalt & 3 &    CONFIRMED &  1.16770  &  PHOT &  0.28916  &    0.086 &  1.08   &    0.102 &    0.7234 &    0.5971 & 1.16402  & 3 \\
6dF J0218081-045845            &  24194304\_1 &  0.71200  & zNEDalt & 3 &     VERIFIED &  2.09593  &  SPEC &  0.91357  &    0.128 &  1.03   &    0.134 &    0.6098 &    0.4552 & 0.70826  & 3 \\
NGC 0720                       &  11083520\_0 &  0.00582  & zNED    & 2 &     PROBABLE &  0.00582  &  SPEC &  0.01562  &    0.162 &  1.32   &    0.222 &    0.3429 &    0.1828 &          &   \\
PKS B1048-238                  &  28146432\_0 &  0.20441  & zIRS    & 3 &     VERIFIED &  0.03216  &  SPEC &  0.20441  &    0.206 &  3.12   &    0.453 &    0.0419 &    0.0066 &          &   \\
SDSS J001342.44-002412.4       &  26906368\_0 &  0.15777  & zIRS    & 3 &     VERIFIED &  1.65050  &  SPEC &  0.15777  &    0.138 &  1.70   &    0.253 &    0.2700 &    0.1252 & 0.15777  & 1 \\
SDSS J160222.38+164353.7       &  17546240\_0 &  0.13825  & zIRS    & 3 &     VERIFIED &  0.67200  & SPEC? &  0.13825  &    0.171 &  1.10   &    0.192 &    0.4244 &    0.2562 &          &   \\
SDSS J161511.06+550625.5       &  11350016\_0 &  0.47500  & zNEDalt & 3 &    CONFIRMED &  1.26500  &   IRS &  1.45705  &    0.087 &  1.11   &    0.109 &    0.6950 &    0.5602 & 0.47327  & 3 \\
SMM J163554.2+661225           &  16210944\_0 &  2.51500  & zNED    & 3 &    CONFIRMED &  2.51500  &  SPEC &  2.52171  &    0.298 &  4.20   &    0.610 &    0.0067 &    0.0004 &          &   \\
SSTXFLS J171538.1+592540       &  11867392\_0 &  2.31497  & zIRS    & 2 &       ROBUST &  2.34000  &   IRS &  2.31497  &    0.082 &  1.04   &    0.090 &    0.7605 &    0.6464 &          &   \\
SSTXFLS J172123.6+595617       &  15523840\_0 &  1.01000  & zIRSeye & 1 &    UNCERTAIN &  1.00000  &  PHOT &  0.98967  &    0.118 &  2.31   &    0.300 &    0.1843 &    0.0685 &          &   \\

\enddata
\tablecomments{This table is available in its entirety in a machine-readable form in the IDEOS website: http://ideos.astro.cornell.edu/redshifts.html}

\end{deluxetable}
\end{landscape}


\begin{thebibliography}{999}

\bibitem[Dasyra et al.(2009)]{Dasyra09}Dasyra K. M. et al., 2009, ApJ, 701, 1123
\bibitem[Farrah et al.(2008)]{Farrah08}Farrah D. et al., 2008, ApJ, 677, 957
\bibitem[Hern\'an-Caballero et al.(2009)]{Hernan-Caballero09}Hern\'an-Caballero A. et al., 2009, MNRAS, 395, 1695
\bibitem[Hern\'an-Caballero \& Hatziminaoglou(2011)]{Hernan-Caballero11}Hern\'an-Caballero A. \& Hatziminaoglou E., 2011, MNRAS, 414, 500
\bibitem[Hern\'an-Caballero(2012)]{Hernan-Caballero12}Hern\'an-Caballero, A., 2012, MNRAS, 427, 816
\bibitem[Houck et al.(2004)]{Houck04}Houck J.R., et al. 2004, SPIE, 5487, 62
\bibitem[Houck et al.(2005)]{Houck05}Houck, J. R., et al. 2005, ApJL, 622, 105
\bibitem[Lebouteiller et al.(2010)]{Lebouteiller10}Lebouteiller, V., Bernard-Salas, J., Sloan, G. C., Barry, D. J. 2010, PASP, 122, 231
\bibitem[Lebouteiller et al.(2011)]{Lebouteiller11}Lebouteiller V., Barry D.J., Spoon H.W.W., Bernard-Salas J., Sloan G.C., Houck J.R., \& Weedman D., 2011, ApJS, 196, 8
\bibitem[Lebouteiller et al.(2015)]{Lebouteiller15}Lebouteiller V., Barry, D. J., Goes, C., Sloan G. C., Spoon, H. W. W., Weedman, D. W., Bernard-Salas, J., Houck, J. R., 2015, ApJSS, 218, 21
\bibitem[Lonsdale et al.(2003)]{Lonsdale03}Lonsdale, C. et al. 2003, PASP 115, 897
\bibitem[Weedman et al.(2006)]{Weedman06}Weedman D. W. et al., 2006a, ApJ, 651, 101
\bibitem[Weedman et al.(2009)]{Weedman09}Weedman D. W. et al., 2009, ApJ, 693, 370
\bibitem[Werner et al.(2004)]{Werner04}Werner, M. W. et al., 2004, ApJS, 154, 1
\bibitem[Yan et al.(2007)]{Yan07}Yan L. et al., 2007, ApJ, 658, 778
	
\end{thebibliography}
\end{document}